\documentclass[a4paper,11pt]{article}
\usepackage{pos}

\title{Modelling uncertainties and prompt b-jet identification in $t\bar{t}b\bar{b}$ production with dilepton signatures at the LHC}
\ShortTitle{Modelling uncertainties and prompt $b$-jet identification in $t\bar{t}b\bar{b}$ at the LHC}

\author*{Giuseppe Bevilacqua}

\affiliation{Institute of Nuclear and Particle Physics, NCSR "Demokritos",\\
  15341 Agia Paraskevi, Greece }

\emailAdd{bevilacqua@inp.demokritos.gr}

\abstract{ As one of the primary backgrounds to $pp \to t\bar{t}H(H\to b\bar{b})$, the QCD process of $t\bar{t}b\bar{b}$ hadroproduction  demands very accurate theoretical predictions and estimates of uncertainties. On top of that, the capacity of properly identifying prompt $b$-jets is beneficial for many analyses. In this contribution we discuss state-of-the-art predictions for $t\bar{t}b\bar{b}$ production with dilepton decays at the LHC, as obtained from full off-shell NLO QCD calculations. We evaluate the dominant uncertainties of the calculation and the impact of the off-shell effects on fiducial cross sections. Furthermore, we explore a kinematic prescription for the categorisation of prompt $b$-jets. }

\FullConference{16th International Symposium on Radiative Corrections: Applications of Quantum Field Theory to Phenomenology (RADCOR2023)\\
28th May - 2nd June, 2023\\
Crieff, Scotland, UK\\}

\begin{document}
\maketitle

\section{Introduction}

With the discovery of the Higgs boson and the first measurements of its properties, the LHC has set milestones in place to refine our understanding of the Electroweak Symmetry Breaking in the Standard Model (SM). To date, all measured properties of the Higgs boson  including couplings, spin and parity are consistent with the SM expectations. One important task for the ongoing LHC Run 3 (and for the future HL-LHC phase as well) is to refine the existing measurements and pave the road to new ones, exploiting both increased statistics and advancements in the analysis techniques. Extensive efforts are being made to study Higgs boson interactions with the highest precision possible \cite{ATLAS:2022vkf,CMS:2022dwd} as new physics beyond the Standard Model (BSM) could manifest in the form of small deviations with respect to SM Higgs couplings up to few percent (see \textit{e.g.} Ref. \cite{Dawson:2022zbb} and references therein).  

As the heaviest elementary particle known so far, the top quark has the largest coupling ($y_t$) to the Higgs boson. The process $pp \to t\bar{t}H(H \to b\bar{b})$ has a great potential among many channels that can be used to measure $y_t$. Its cross section has direct sensitivity to the coupling and benefits from the large branching fraction of the $H \to b\bar{b}$ decay. The limiting factor for precision studies is the presence of huge backgrounds. Particularly insidious is the \textit{irreducible} QCD background $pp \to t\bar{t}b\bar{b}$, which shares the same final-state composition with the signal. The latter background dominates final states with at least four tagged $b$-jets \cite{CMS:2018hnq} and demands accurate theoretical modelling.

The $t\bar{t}b\bar{b}$ process has been studied extensively over the past decade. Measurements of the $t\bar{t}b\bar{b}$ cross sections at $\sqrt{s} = 13$ TeV have been reported by the CMS and ATLAS collaborations \cite{CMS:2018hnq,CMS:2017xnm,ATLAS:2018fwl,CMS:2020grm}.
From the theory side, state-of-the-art accuracy is NLO in QCD. The process has been studied at fixed-order  \cite{Bredenstein:2009aj,Bevilacqua:2009zn,Bredenstein:2010rs,Bevilacqua:2014qfa,Buccioni:2019plc,Denner:2020orv,Bevilacqua:2021cit,Bevilacqua:2022twl} as well as in the context of NLO+PS simulations  \cite{Kardos:2013vxa,Garzelli:2014aba,Cascioli:2013era,Bevilacqua:2017cru,Jezo:2018yaf}. Overall, estimates of cross sections suffer from large uncertainties in the choice of factorisation and renormalisation scales that are of the order of $20\%-30\%$. The size of QCD corrections is also quite large and can reach the order of 80\% depending on the choice of scales. These facts suggest that contributions from missing higher orders might be important for this process.

In this contribution we report on state-of-the-art predictions for $t\bar{t}b\bar{b}$ production with di-lepton signatures, as presented in \cite{Bevilacqua:2021cit,Bevilacqua:2022twl}. The emphasis is on estimates of fiducial cross sections and the associated uncertainties. By comparing results obtained with matrix elements at different levels of accuracy, we discuss effects such as the impact of QCD corrections to top-quark decays and the size of the off-shell effects on the NLO calculation. We show two examples of kinematic observables which are particularly sensitive to off-shell effects. Finally, we describe a simple kinematic prescription to categorise prompt $b$-jets. The latter could be also beneficial for the analysis of the $t\bar{t}H(H \to b\bar{b})$ signal.

\section{Setup of the calculation}

\begin{figure}[h!tb]
\centerline{
\put(-210,0){\includegraphics[width=0.3\textwidth]{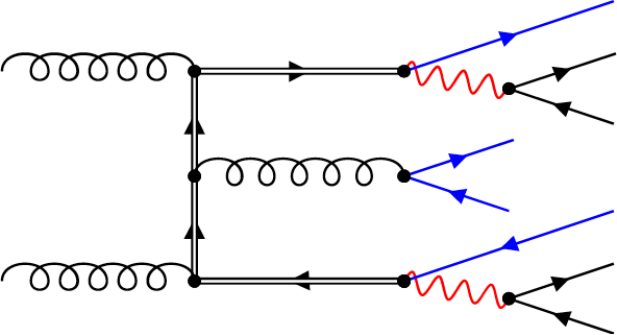}}
\put(-152,-13){($a$)}
\put(-63,8){\includegraphics[width=0.3\textwidth]{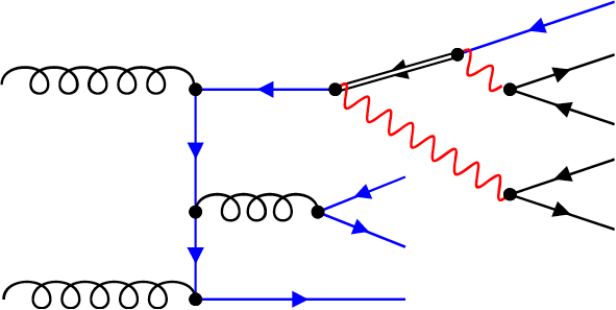}}
\put(-3,-13){($b$)}
\put(80,8){\includegraphics[width=0.3\textwidth]{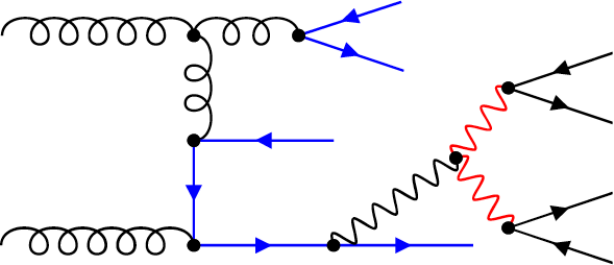}}
\put(140,-13){($c$)}
}
\caption{Representative Feynman diagrams contributing to the process $gg \to e^+ \nu_e \, \mu^- \bar{\nu}_ \mu \, b \bar{b} \, b \bar{b} $ at leading order: \textit{double-resonant} ($a$), \textit{single-resonant} ($b$), \textit{non-resonant} ($c$). The double-lined propagators represent top quarks, the red propagators $W$ bosons and the blue lines bottom quarks.}
\label{Fig:diagrams}
\end{figure}
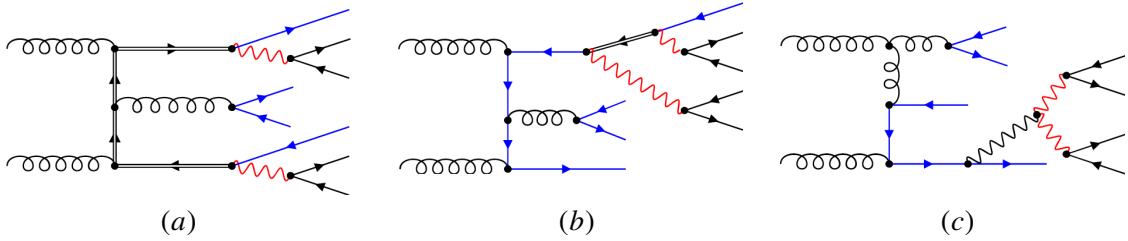

We consider the process $pp \to e^+\nu_e \, \mu^-\bar{\nu}_\mu \, b\bar{b} \, b\bar{b} + X$ at the collider energy $\sqrt{s}=13$ TeV. 
The perturbative accuracy is NLO in QCD. Our reference predictions are based on a full off-shell calculation, namely a complete calculation of matrix elements for the given final state at the order $\mathcal{O}(\alpha^4 \alpha_s^3)$. The off-shell calculation considers all double-, single- and non-resonant contributions to matrix elements (see Figure \ref{Fig:diagrams}) and interferences among them. We compare it against the Narrow Width Approximation (NWA), based on the limits $\Gamma_t/m_t \to 0$ and $\Gamma_W/m_W \to 0$. The NWA restricts the calculation to double-resonant contributions only. Spin-correlation effects in decays of top quarks and $W$ bosons are taken into account. Moreover, QCD corrections are incorporated in production and decay subprocesses  consistently with the expansion in the strong coupling constant.

The details of the analysis are briefly summarised. We require two charged leptons and at least four $b$-jets in the final state. Jets are defined using the anti-$k_T$ algorithm \cite{Cacciari:2008gp} with resolution parameter $R=0.4$. The following kinematical cuts are imposed on charged leptons ($\ell = e,\mu$) and $b$-jets:
\begin{equation}
p_T(\ell) > 20 \,{\rm GeV}      \,,\qquad         p_T(b) > 25 \,{\rm GeV}           \,,\qquad  
\vert y(\ell) \vert < 2.5     \,,\qquad        \vert y(b) \vert < 2.5  \,.
\end{equation}
No kinematical restriction is applied to the missing transverse momentum nor to the extra jet.
The reference choice for the renormalisation and factorisation scale is
\begin{equation}
\mu_R = \mu_F = H_T/3  \,,\qquad H_T = \sum_{i=1}^{4} p_{T,b_i} + \sum_{i=1}^{2} p_{T,\ell_i} + p_{T,miss} \,,
\end{equation}
where $b_i \,(i=1,4)$ denotes the four hardest resolved $b$-jets in the final state. Scale uncertainties are estimated using standard 7-point variation.
Further details, including the numerical parameters used for the calculation, can be found in Ref. \cite{Bevilacqua:2022twl}.

On the technical side, all results have been obtained with the help of the package \textsc{Helac-Nlo} \cite{Bevilacqua:2011xh}, which consists of \textsc{Helac-1Loop} \cite{vanHameren:2009dr} and \textsc{Helac-Dipoles} \cite{Czakon:2009ss, Bevilacqua:2013iha}, and uses   \textsc{CutTools} \cite{Ossola:2007ax}, \textsc{OneLOop} \cite{vanHameren:2010cp} and \textsc{Kaleu} \cite{vanHameren:2010gg} as its cornerstones. 
The framework has been applied already to get predictions for a variety of processes describing associated top-quark production, see \textit{e.g.} Refs. \cite{Bevilacqua:2016jfk,Bevilacqua:2018woc,Bevilacqua:2019cvp,Bevilacqua:2019quz,Bevilacqua:2020pzy,Bevilacqua:2022ozv}.
Predictions are stored in the form of modified Les Houches Event files \cite{Alwall:2006yp} and ROOT Ntuples \cite{Antcheva:2009zz}, where each event is equipped with additional matrix-element and PDF information \cite{Bern:2013zja}. Uncertainties stemming from scale and PDF variation are determined reweighting the events. This allows us to obtain fairly accurate uncertainty estimates without need to perform multiple runs of the code.

\section{Phenomenological results}

\begin{figure}[h!tb]
\centerline{
\includegraphics[width=0.6\textwidth]{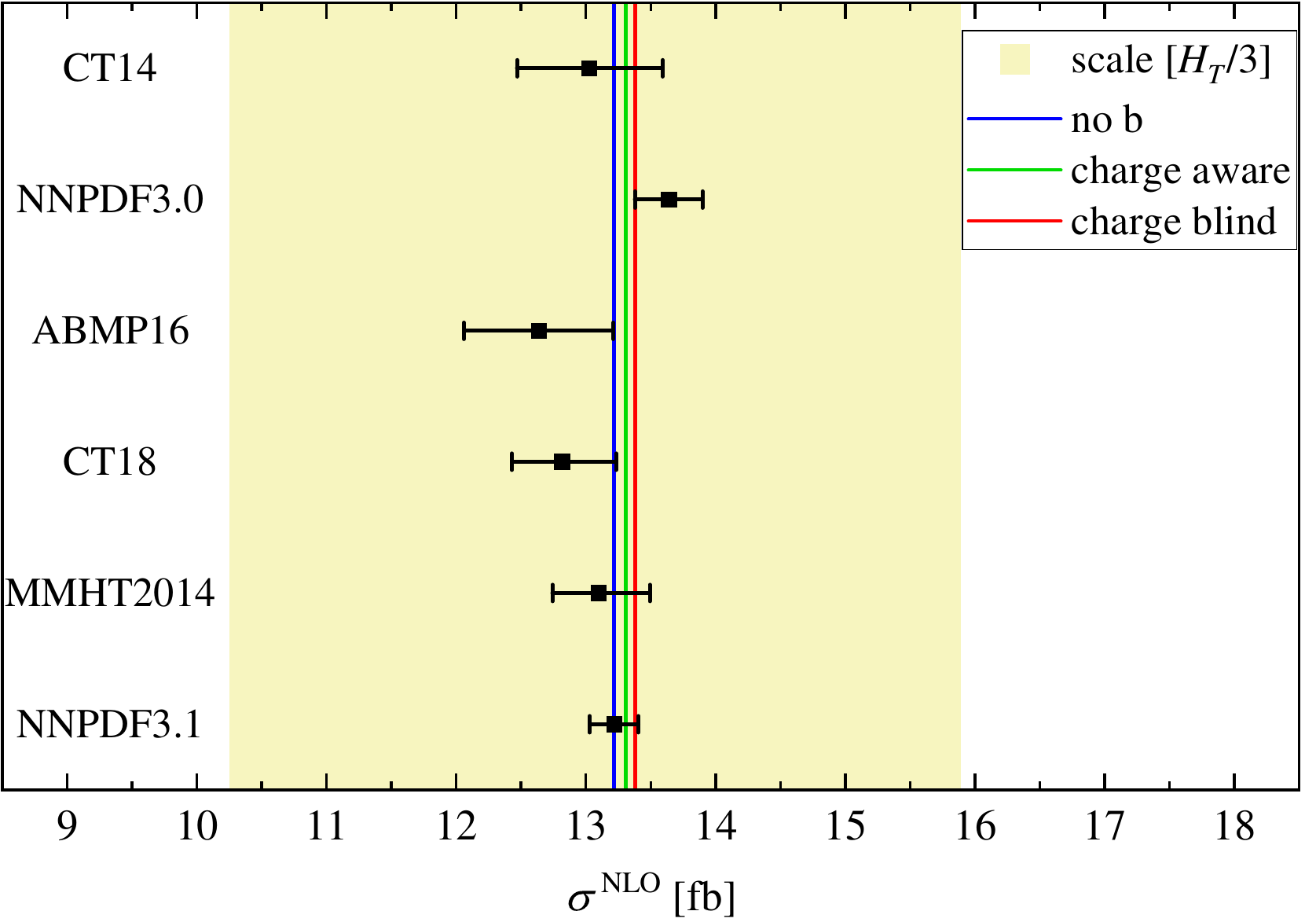}
}
\caption{NLO fiducial cross section for the process $pp \to e^+\nu_e \mu^-\bar{\nu}_\mu b\bar{b}b\bar{b} + X$ at the LHC with $\sqrt{s} = 13$ TeV. The vertical lines indicate the reference cross section based on \textsc{NNPDF3.1} PDF set. The yellow band represents scale uncertainties for $\mu_0 = H_T/3$, while the black error bars denote internal PDF uncertainties. Figure taken from Ref. \cite{Bevilacqua:2021cit}.}
\label{Fig:theory_uncert}
\end{figure}

We begin the discussion with a look at the integrated fiducial cross section and the associated theory uncertainties. Figure \ref{Fig:theory_uncert} presents a synoptic view of the main uncertainties affecting our calculation. As typical of NLO QCD predictions, the dominant uncertainty is related to scale dependence of the cross section (represented by the yellow band) and amounts to $22\%$. Less sizeable are the uncertainties connected to PDF variation (shown as black error bars). We have considered several PDF sets: NNPDF3.1 \cite{NNPDF:2017mvq} (our default choice), MMHT2014 \cite{Harland-Lang:2014zoa}, CT18 \cite{Hou:2019efy}. The latter are part of the PDF4LHC recommendations for applications at the LHC Run 2 \cite{Butterworth:2015oua}. On top of them we show also CT14 \cite{Dulat:2015mca}, NNPDF3.0 \cite{Ball:2014uwa} and ABMP16 \cite{Alekhin:2018pai} for comparison. Overall, PDF uncertainties are set in the ballpark of $3\%$. As a bonus of our study, we have also investigated the impact of contributions from $b$-quark initiated subprocesses. The latter are expected to be suppressed by small $b(\bar{b})$-parton luminosities: we find indeed that they contribute to the NLO cross section up to 1\%. Thus, they can be safely neglected given the dominant uncertainties. For details concerning $b$-jet definitions (charge-aware and charge-blind tagging), we point to Ref. \cite{Bevilacqua:2021cit}.

\begin{figure}[h!tb]
\centerline{
\includegraphics[clip, trim=0.cm 6.5cm 0.cm 6.5cm, width=0.9\textwidth]{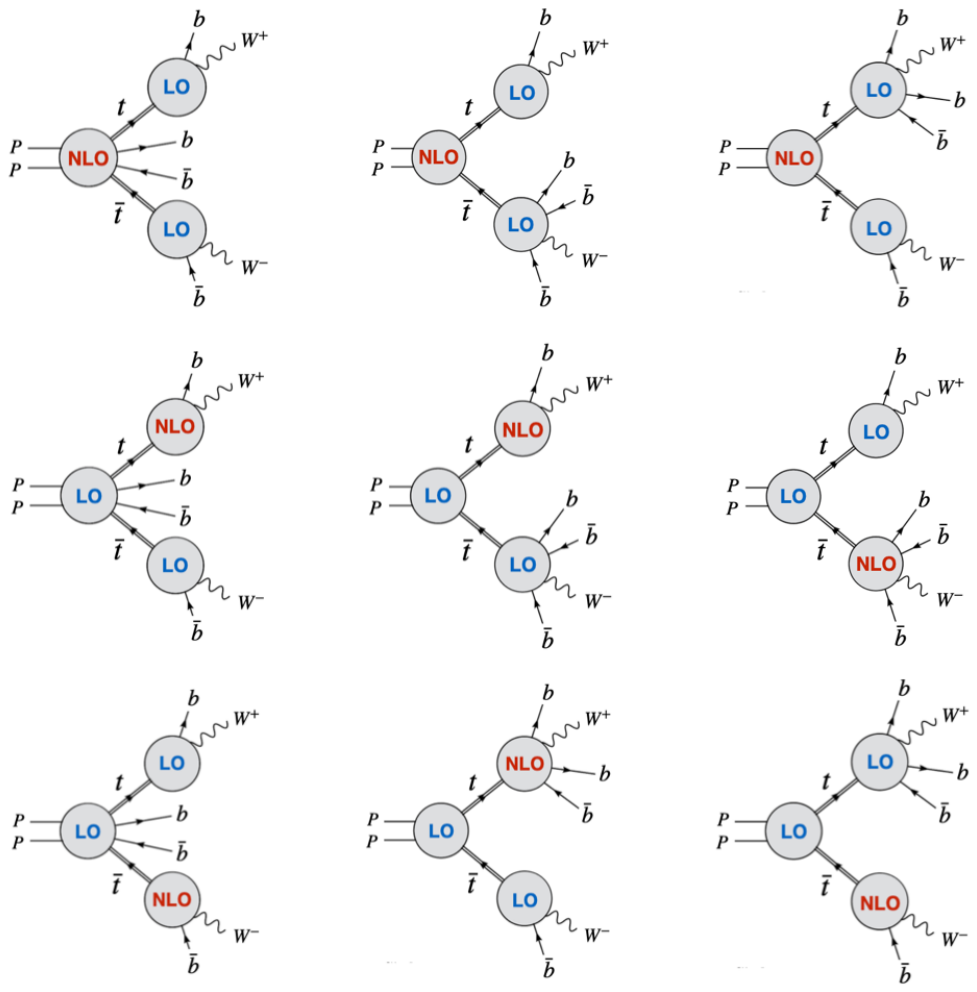}
}
\caption{Resonant structures contributing to the NLO cross section of the process $pp \to e^+\nu_e \mu^-\bar{\nu}_\mu b\bar{b}b\bar{b} + X$ in full NWA. The blobs denote schematically matrix elements describing \textit{production} and \textit{decay} subprocesses.}
\label{Fig:resonant_contrib}
\end{figure}

In Table \ref{Tab:ttbb_xsec} we compare fiducial cross sections obtained using different levels of accuracy in the calculation of matrix elements. As already mentioned, our reference prediction is the full off-shell calculation and the NWA is used for comparison. 
We would like to stress that, generally speaking, the NWA should not be identified as naive "\textit{$t\bar{t}b\bar{b}$ production $\times$ decay}". The reason is that the $b\bar{b}$ pair can be either part of the production subprocess ($pp \to t \bar{t} b \bar{b} \to W^+ W^- b \bar{b} b \bar{b}$) or the product of a top-quark decay ($pp \to t \bar{t} \to W^+ W^- b \bar{b} b \bar{b}$). These two different kinds of production dynamics (that we refer as \textit{resonant structures} in the following) are non-interfering in NWA and contribute both to the full cross section ($\rm NWA_{full}$). They are illustrated schematically in Figure \ref{Fig:resonant_contrib}. 
The impact of genuine off-shell effects at the integrated level can be inferred by comparing the first two rows in Table \ref{Tab:ttbb_xsec}: we find that they amount to $0.5\%$. The other rows report partial results, as obtained by selecting only specific resonant structures out of the full set defining $\mathrm{NWA_{full}}$. To be more precise, $\mathrm{NWA_{LOdec,prod}}$ takes into account only the first element shown in Figure \ref{Fig:resonant_contrib}, whereas $\mathrm{NWA_{LOdec}}$ ($\mathrm{NWA_{prod}}$) consider only elements appearing in the first row (column). These results help us to infer the numerical impact of $(i)$ contributions from multi-particle decays ($t \to W^+ b b \bar{b}$ or $\bar{t} \to W^- \bar{b} b \bar{b}$) and $(ii)$ QCD corrections to decay subprocesses. Concerning point $(i)$, our conclusions are that effects from multi-particle decays are negligible in our analysis (indeed $\mathrm{NWA_{prod}}$ differs with $\mathrm{NWA_{full}}$ by only $1.1\%$).  This is not surprising, as the $b$-jet kinematical cuts suppress the phase space of on-shell $t \to W b b \bar{b}$ decays to a larger extent compared to $t \to W b$ decays. Of course, this kind of effects is expected to increase in absolute size when looser kinematical cuts are considered. Concerning point $(ii)$, using LO decays in the calculation gives subpercent agreement with the full NWA result (as it is clear by comparing $\mathrm{NWA_{LOdec}}$ and $\mathrm{NWA_{full}}$). It is tempting to conclude that the impact of QCD corrections to decays is quite small. However, the proper interpretation of this result requires to identify and disentangle two different effects. It is instructive to compare $\mathrm{NWA_{prod}}$ and $\mathrm{NWA_{prod,exp}}$ in Table \ref{Tab:ttbb_xsec}: the latter is obtained from the former after consistent expansion in $\alpha_S$ of the NLO top-quark width parameter (see Ref. \cite{Bevilacqua:2022twl} for further details). The observed difference, of the order of $6\%$, is understood as formally suppressed higher-order contributions (\textit{i.e.}, $\mathcal{O}(\alpha_S^2)$ with respect to the Born level) which enter the unexpanded NWA cross section\footnote{The unusually large size of these contributions is correlated to the large $K$-factor affecting the $t\bar{t}b\bar{b}$ production cross section. We note that these effects are well within the estimated scale uncertainties of the order of $22\%$.}. The genuine impact of QCD corrections to top decays becomes evident when the $\mathcal{O}(\alpha_S^2)$ contributions are systematically removed, and it amounts to approximately $-7\%$. The quite strong cancellation between these two effects is accidental.
\begin{table}[t]
\centering
\begin{tabular}{llll}
\hline \hline \\[-0.44cm]
  Modelling &$\sigma^{\rm NLO}$ [fb] & $\delta_{\rm scale}$ [fb]  & $\frac{\sigma^{\rm NLO}}{\sigma^{\rm NLO}_{\rm {NWA_{full}}}}-1$  \\[0.3cm]
 \hline \hline \\[-0.41cm]
 Off-shell & $13.22(2)$   & ${}^{+2.65~(20\%)}_{-2.96~(22\%)}$ & $+0.5\%$ \\[0.15cm]
\hline  \\[-0.41cm]
 $\rm NWA_{full}$ & $13.16(1)$ & ${}^{+2.61~(20\%)}_{-2.93~(22\%)}$  &  $-$ \\[0.15cm]
 \hline \\[-0.41cm]
 $\rm NWA_{ LOdec}$ & $13.22(1)$ & ${}^{+3.77~(29\%)}_{-3.31~(25\%)}$  & $+0.5\%$ \\[0.15cm]
 \hline \\[-0.41cm]
 $\rm NWA_{ prod}$ & $13.01(1)$ & ${}^{+2.58~(20\%)}_{-2.89~(22\%)}$  & $-1.1\%$ \\[0.15cm]
 \hline \\[-0.41cm]
 $\rm NWA_{ LOdec, prod}$ & $13.11(1)$ & ${}^{+3.74~(29\%)}_{-3.28~(25\%)}$ & $-0.4\%$ \\ [0.15cm]
\hline \hline \\[-0.41cm]
 $\rm NWA_{ prod,exp}$ & $12.25(1)$ & ${}^{+2.87~(23\%)}_{-2.86~(23\%)}$  & $-6.9\%$ \\[0.15cm]
\hline \hline
\end{tabular}
\caption{NLO fiducial cross sections for the process $pp \to e^+\nu_e \mu^-\bar{\nu}_\mu b\bar{b}b\bar{b} + X$ at $\sqrt{s}= 13$ TeV, based on different top-quark decay modelling accuracies \cite{Bevilacqua:2022twl}. The reference choices for scale and PDF set are $\mu_R = \mu_F = H_T/3$ and NNPDF3.1 respectively.}
\label{Tab:ttbb_xsec}
\end{table}

\begin{figure}[h!tb]
\centerline{
\includegraphics[clip, trim=0.cm 4.5cm 0.cm 5.cm, width=0.5\textwidth]{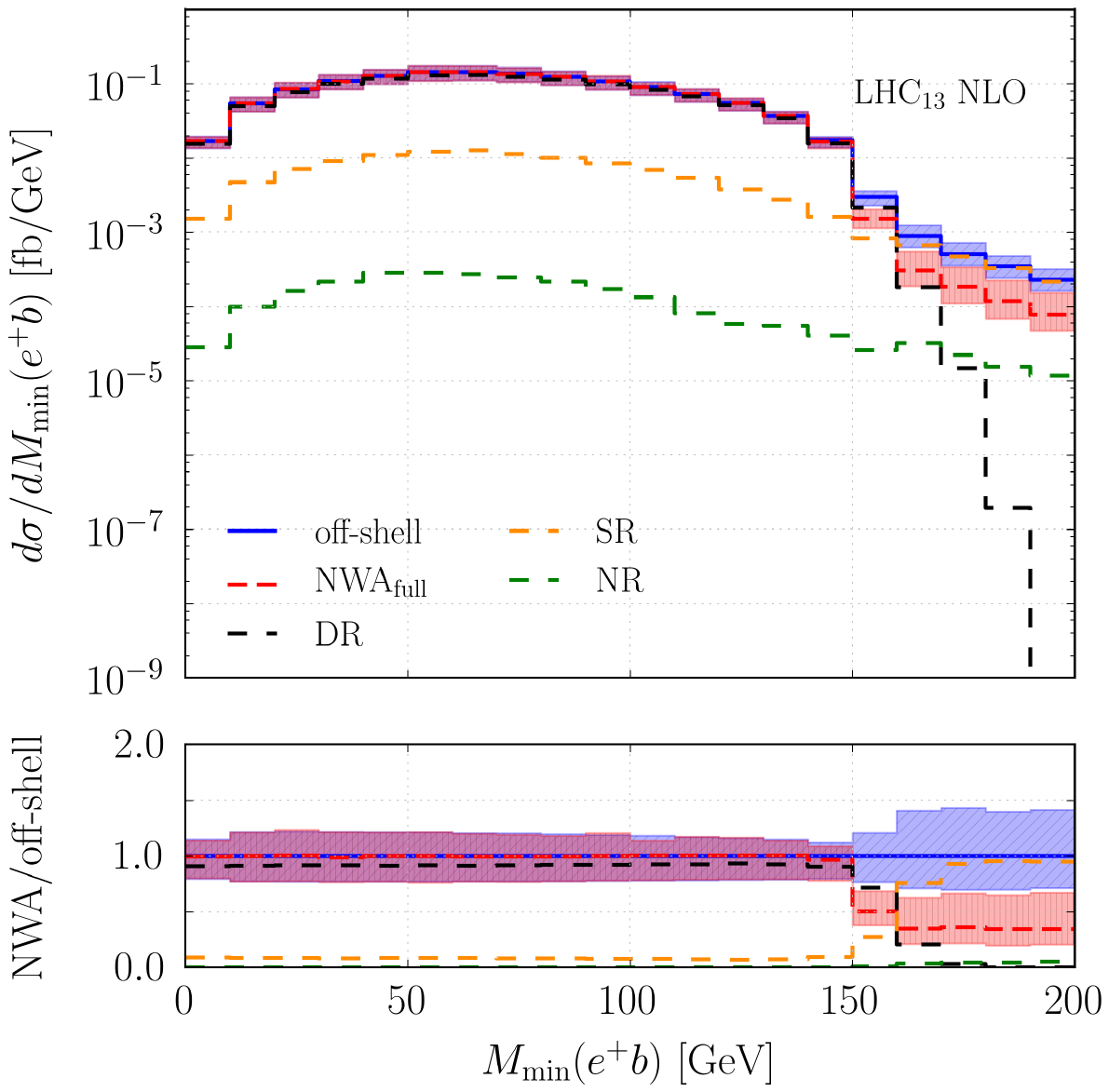}
\includegraphics[clip, trim=0.cm 4.5cm 0.cm 5.cm, width=0.5\textwidth]{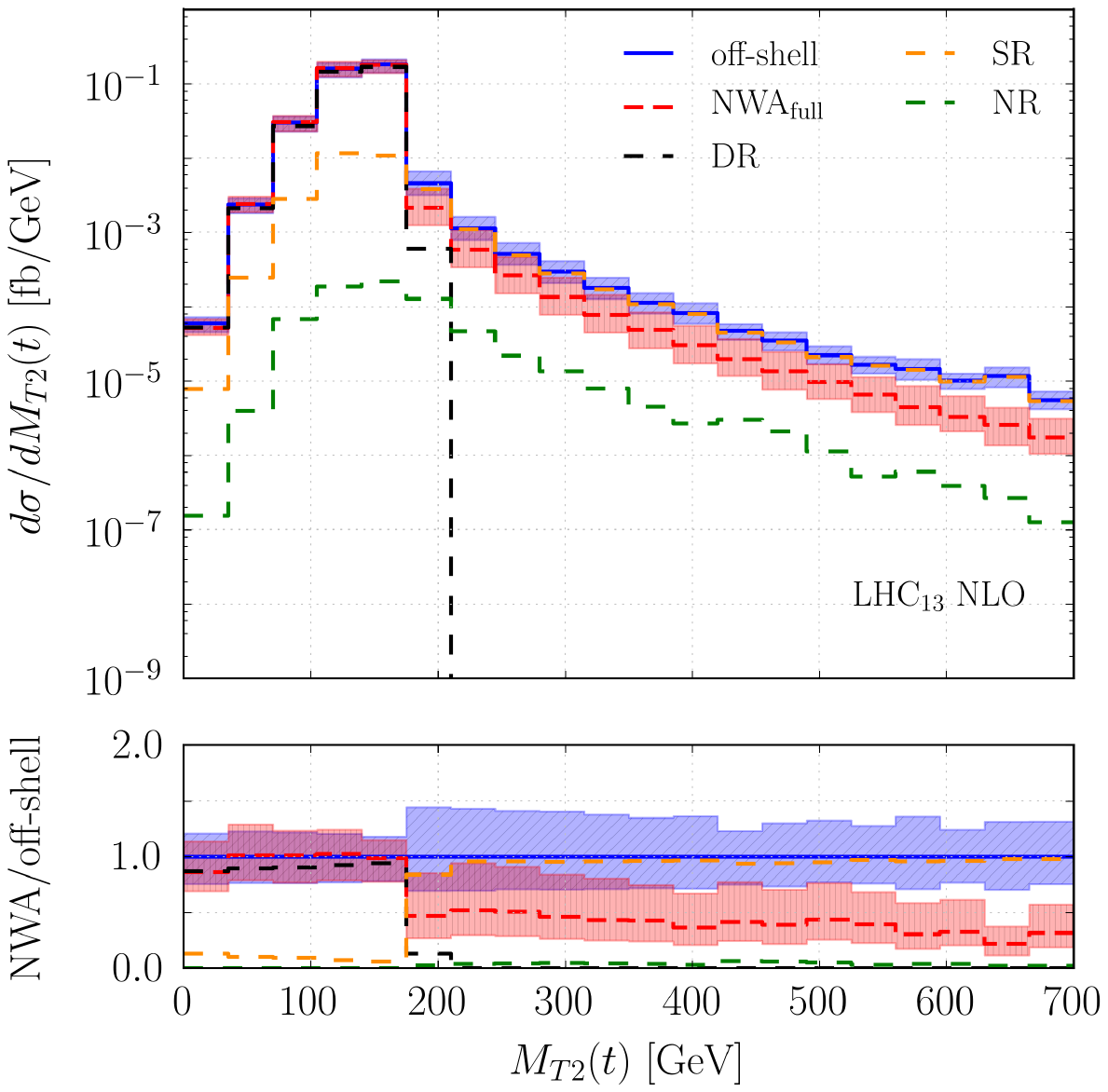}
}
\caption{Differential cross section as a function of $M_{min}(e^+b)$ (\textit{left plot}) and $M_{T2}(t)$ (\textit{right plot}). The observables are defined in the text. Figures taken from Ref. \cite{Bevilacqua:2022twl}}
\label{Fig:offshell_diff}
\end{figure}

In the next step, we make some considerations about the impact of the off-shell effects at differential level. We have scrutinised a large set of observables typically considered in SM analyses, both dimensionful (\textit{e.g.}: $M(b\bar{b})$, $p_T(b\bar{b})$, $M(e^+\mu^-)$, $H_T$, etc.) and dimensionless (\textit{e.g.}: $\Delta R(b\bar{b})$, $\Delta\phi(e^+\mu^-)$, etc.). The off-shell effects are found to be at subpercent level in the whole observed range. Thus, the NWA does a very good job in all these cases. Where the NWA shows its weaknesses is the modelling of \textit{threshold observables}, which are most interesting for BSM analyses. In Figure \ref{Fig:offshell_diff} we show two examples of them, namely $M_{min}(e^+b)$ and $M_{T2}(t)$.  The latter is defined as \cite{Lester:1999tx}
\begin{equation}
 M_{T2}(t) \: = \min_{\sum_i p_T^{\nu_i} = p_t^{miss}}  \left[ \, \max \{ \, M_T\left( p_T(e^+\,X_t), p_T(\nu_1) \right),  M_T\left( p_T(\mu^-\,X_{\bar{t}}), p_T(\nu_2) \right)  \}  \right] 
\end{equation}
and can be understood as a proxy to top-quark transverse mass. $M_{min}(e^+b)$ is also well known for being used in top-quark mass measurements in the context of $t\bar{t}$ production by the CMS and ATLAS collaborations \cite{ATLAS:2016muw,CMS:2017znf}. Both observables have been used in BSM analyses, for example to disentangle possible new heavy resonances from the QCD background (see \textit{e.g.} Ref. \cite{Haisch:2018bby,Hermann:2021xvs,ATLAS:2014gmw}). It is not difficult to see that these observables exhibit a kinematical edge around $M_{min}(e^+b) = \sqrt{m_t^2 - m_W^2} \approx 153$ GeV and $M_{T2}(t) = mt = 173$ GeV. The edge can be explained as a sharp kinematical threshold (characteristic of the NWA prediction at LO) which is smeared out by off-shell effects and (starting from NLO) by extra QCD radiation. One can see that, while the $\mathrm{NWA_{full}}$ prediction looks fairly accurate in the region below the kinematical edge, it badly underestimates the full off-shell result beyond it. Thus, a precise modelling in the full kinematical range can only be achieved using full off-shell calculations.

\begin{figure}[h!tb]
\centerline{
\includegraphics[clip, trim=0.cm 5.8cm 0.cm 4.7cm, width=0.5\textwidth,height=6.3cm]{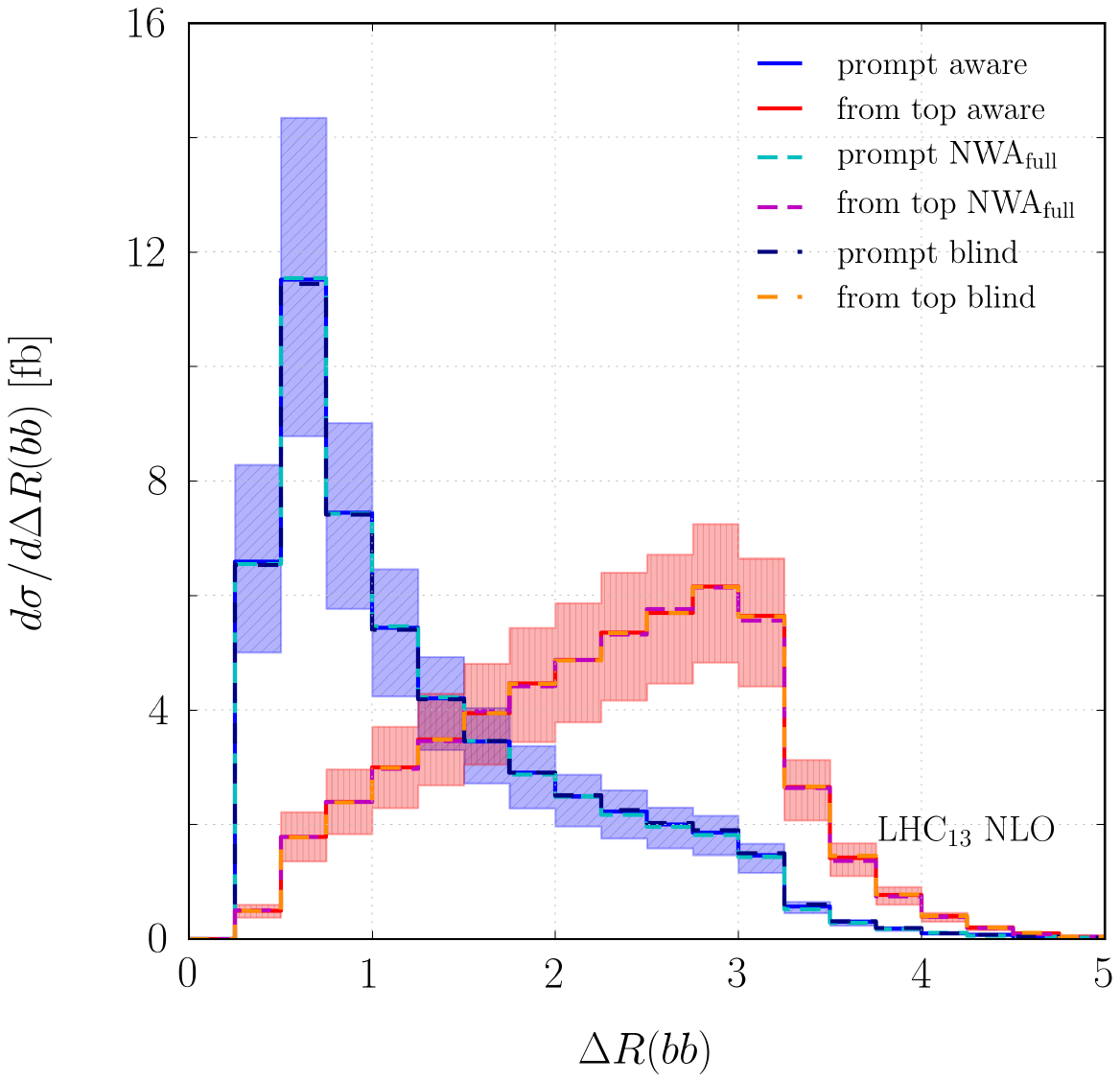}
\includegraphics[clip, trim=0.cm 5.8cm 0.cm 4.7cm, width=0.5\textwidth,height=6.3cm]{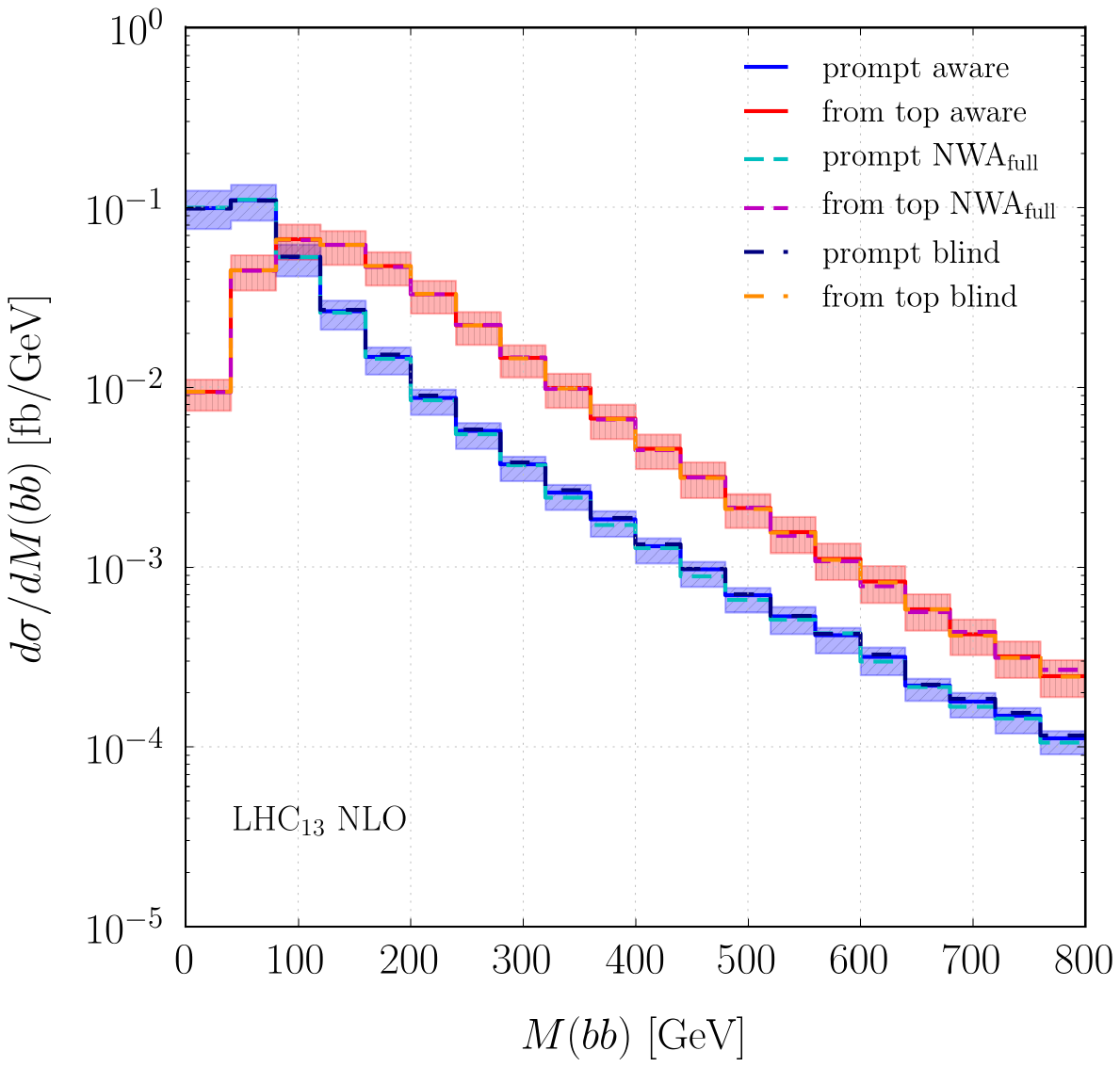}
}
\caption{Kinematics of \textit{prompt} (blue) versus \textit{non-prompt} (red) $b$-jets in function of $\Delta R$ separation (\textit{left plot}) and invariant mass (\textit{right plot}). The identification of prompt $b$-jets is performed according to the prescription described in the text. Figures taken from Ref. \cite{Bevilacqua:2022twl}}
\label{Fig:bjet_label}
\end{figure}

In the last part of the discussion, we address the categorisation of prompt $b$-jets. The word "prompt" denotes ideally the $b$-jet pair  produced in association with top quarks (\textit{i.e.}, not originated by decays). Identifying these jets properly is important for  $t\bar{t}H(H\to b\bar{b})$ measurements, as it impacts the reconstruction of the Higgs boson. It should be clear that such identification is not free of ambiguities conceptually. As we have seen, $t\bar{t}b\bar{b}$ is encompassed in double-resonant diagrams (Figure \ref{Fig:diagrams}, $(a)$). Altogether the latter form a sub-amplitude, not even gauge invariant, which interferes with other sub-amplitudes entering the full matrix element. Even when the simplified picture of NWA is considered, the definition of prompt $b$-jets is made ambiguous by the presence of $pp \to t\bar{t} \to W^+ W^- b \bar{b} b \bar{b}$ resonant structures. However, since we have checked that the latter are numerically small, this is not a serious issue in practice. The $\mathrm{NWA_{full}}$ result can be used as "approximate Monte Carlo truth" to assess the validity of phenomenological prescriptions that are developed to categorise $b$-jets. The prescription that we adopt in our study can be summarised in the following steps:
\begin{enumerate}
\item for a given event, we consider all possible ways to assign the $b$-jets in the final state to the decay of a top quark ($t$), or an anti-top quark ($\bar{t}$), or to none of them;
\item for each assignment, the momenta of $t$ and $\bar{t}$ are defined using information from candidate daughters. The $b$-jets that have not been assigned to any decay are considered prompt candidates ($b\bar{b}_{\rm prompt}$). If all $b$-jets have been assigned to decays, we define $b\bar{b}_{\rm prompt}$ as the two jets with smallest $p_T$;
\item the following quantity is computed for each assignment:
\begin{equation}
Q = \vert M(t) - m_t \vert \, \times \, \vert M(\bar{t}) - m_t \vert \, \times \,  M(b\bar{b}_{\rm prompt})  \,;
\end{equation}
\item we select the assignment which minimises the $Q$ function and identify prompt $b$-jets accordingly.
\end{enumerate}
Figure \ref{Fig:bjet_label} shows the performance of our prescription when applied to the full off-shell calculation. We consider two case studies, namely $\Delta R(b\bar{b})$ and $M(b\bar{b})$ distributions. The different kinematics of prompt $b$-jets as opposed to jets associated with top decays (represented in blue and red respectively) is evident. We note that prompt $b$-jet distributions have their bulk at smaller $\Delta R(b\bar{b})$ and $M(b\bar{b})$ values. This is consistent with the fact that the prompt $b\bar{b}$ pair, in QCD $t\bar{t}b\bar{b}$ production, is originated by $g \to b\bar{b}$ splitting. In all cases we observe excellent agreement with the corresponding result based on NWA. We also note that our results look consistent with alternative procedures based on deep neural network techniques \cite{Jang:2021eph}. All these observations provide evidence that our prescription for the categorisation of $b$-jets works properly.

%

\section{Summary}

We have presented state-of-the-art predictions for $pp \to t\bar{t}b\bar{b}$ production with di-lepton decays at $\sqrt{s} = 13$ TeV, based on a full off-shell calculation at NLO QCD accuracy. Overall, scale dependence is the dominant theory uncertainty and affects fiducial cross sections at the level of $22\%$, while PDF uncertainties are at the order of $3\%$. The impact of the off-shell effects has been studied at integrated and differential level. We have discussed a few examples of observables which require full off-shell calculations for proper modelling. Finally, we have proposed a simple kinematic procedure to categorise prompt $b$-jets.

\acknowledgments{
The research of G.B. is supported by the Hellenic Foundation for Research and Innovation (H.F.R.I.) under the "2nd Call for H.F.R.I. Research Projects to support Faculty Members \& Researchers" (Project Number: 02674 HOCTools-II).
}

\end{document}